\documentclass[aps,pra,onecolumn,floatfix]{revtex4}

\usepackage{graphicx}
\usepackage{amsmath,amssymb}
\usepackage{graphpap}
\usepackage{color}
\usepackage{subfig}

\begin{document}

\title{Lifetimes of Metal Nanowires with Broken Axial Symmetry}

\author{Lan Gong}
\affiliation{Department\ of Physics,~New York University,~New York,~NY 10003}

\author{J.\ B{\"u}rki}
\affiliation{Department of Physics and Astronomy,~California State University Sacramento,~Sacramento,~CA 95819-6041}

\author{Charles A.\ Stafford} 
\affiliation{Department of Physics, University of Arizona, 1118 E.\ 4th St., Tucson, AZ, 85721}

\author{Daniel L.\ Stein}
\affiliation{Department\ of Physics,~New York University,~New York,~NY 10003}
\affiliation{Courant Institute of Mathematical Sciences,~New York University,~New York,~NY 10003}

\date{\today}

\begin{abstract}
  We present a theoretical approach for understanding the stability of
  simple metal nanowires, in particular monovalent metals such as
  the alkalis and noble metals. Their cross sections are of order one
  nanometer so that small perturbations from external (usually
  thermal) noise can cause large geometrical deformations. The
  nanowire lifetime is defined as the time required for making a
  transition into a state with a different cross-sectional
  geometry. This can be a simple overall change in radius, or a change
  in the cross section shape, or both. We develop a stochastic field
  theoretical model to describe this noise-induced transition process,
  in which the initial and final states correspond to locally stable
  states on a potential surface derived 
by solving the Schr\"odinger equation for the electronic structure of the nanowire numerically.
The numerical {\em string method} is implemented to
  determine the optimal transition path governing the lifetime. Using
  these results, we tabulate the lifetimes of sodium and gold
  nanowires for several different initial geometries.
\end{abstract}


\maketitle

\section{Introduction}
\label{sec:intro}

Nanowires made of monovalent metals, such as sodium, copper, and gold,
live at the boundary between classical and quantum mechanics and
exhibit some of the behavior of each.  They are therefore of great
interest from both a fundamental physics perspective as well as for 
technological applications. Their cross-sectional dimensions can be as
small as half a nanometer (though several nanometers is far more
typical) and their lengths at most a few tens of nanometers.  At these
length scales, the stresses induced by surface tension exceed 
Young's modulus, making the wires subject to deformation under plastic
flow~\cite{BS05}, and therefore subject to breakup due to the Rayleigh
instability~\cite{Plateau1873,Burki03}. This in fact has been observed
for copper nanowires annealed between 400 and 600$^\circ$C~\cite{MBCNT04},
as well as for copper and silver nanowires~\cite{Bid05,Karim06}.

However, 
electron-shell filling effects stabilize these wires for radii near 
certain discrete ``magic radii''~\cite{BS05}. These radii correspond
to conductance ``magic numbers'' that agree with those measured in
experiments~\cite{YYR99,YYR00,UBZSG04,Urban04,UBSG06}. This quantum stabilization, however, is only
against small surface oscillations that can lead to breakup via the
Rayleigh instability; it does not take into account thermal noise that
can induce large radial fluctuations that can lead to breakup. 

A self-consistent approach to determining lifetimes~\cite{BSS05},
which modeled thermal fluctuations through a stochastic
Ginzburg-Landau classical field theory, obtained quantitative
estimates of alkali nanowire lifetimes \cite{BSS04,BSS05} in good agreement with
experimentally inferred values~\cite{YYR99,YYR00}. The theory, however,
is limited to wires with a cylindrical symmetry.  
Urban~et al.~\cite{UBZSG04,UBSG06}, using a stability analysis of
metal nanowires subject to non-axisymmetric perturbations, showed
that, at certain mean radii and aspect ratios, Jahn-Teller
deformations breaking cylindrical symmetry can be energetically
favorable, leading to an additional class of stable nanowires with
{\it non\/}-axisymmetric cross sections.

The mathematical problem of determining nanowire lifetimes in this
more general case requires solution of a stochastic set of coupled
partial differential equations corresponding to a stochastic
Ginzburg-Landau field theory with {\it two\/} coupled fields, one
corresponding to variations in mean radius and the other to
deviations from axial symmetry.
In particular, we consider here quadrupolar deformations of the nanowire cross-section, which
cost less surface energy than higher-multipole deformations, and were shown to be the most common stable
deformations within linear stability analyses \cite{UBSG06,Mares07}.
The general mathematical treatment of such
problems was discussed in~\cite{GS10,GS11}. Of particular interest was
the discovery of a transition in activation behavior not only as wire
length varies~\cite{BSS05}, but also as bending coefficients for the
two fields vary~\cite{GS11}. 

In this paper, we use these results and those of Urban et al.~\cite{UBZSG04,UBSG06}
to construct a more general theory of lifetimes of nanowires with both
axisymmetric and nonaxisymmetric cross-sections, as functions of
temperature, strain, and other thermodynamic variables.

\section{Overview of the nanowire stability problem}
\label{sec:exps}
Historically, there have been a number of studies on nanowires focusing on
aspects ranging from growth techniques to electronic,
mechanical, thermal and optical properties. 
There are several major
laboratory techniques for synthesizing nanowires: suspension,
vapor-liquid-solid (VLS), solution-based growth, and so on. The materials
used to fabricate nanowires also vary, including metals (e.g., Ni, Pt,
Au), semiconductors (e.g., Si, {In}{P}, {Ga}{N}, etc.), and insulators (e.g., {Si}{O}$_2$,
{Ti}{O}$_2$). Usually, the focus is on nanowires with cross sections of the order
of hundreds of nanometers or even a few micrometers~\cite{IS06, Astafiev12,Schmid11}. 
However, the type of nanowire under study here is much smaller---at 
most a few nanometers in radius \cite{Agrait03}. 
The nanowires we study are mainly prepared via the suspension technique in
either vacuum or air, as opposed to wires fabricated on a substrate and
adhering to a surface.

Metal nanowires of atomic cross section are typically prepared
using either a scanning tunneling microscope (STM) or a mechanically-controllable 
break junction (MCBJ) \cite{Agrait03}.  In both cases, the nanowires are
essentially freely suspended three-dimensional wires.
In the STM setup, a nanowire is obtained by pushing the sharp STM tip into a
substrate and then carefully retracting it in a controlled way. The contact
formed in this way is of atomic dimensions~\cite{ARV93}. If the tip is
moved further away, the contact thins down and eventually breaks. 
In the MCBJ method~\cite{Ruiten92pc,Ruiten92prl}, the sample is fixed
onto an insulating substrate and then bent by a piezoelectric drive controlled by
an applied voltage. The sample is pulled apart until the wire breaks,
after which it reforms by reversing the process. The displacement
between the electrodes can be controlled with accuracy down to
100fm. Because this setup is more stable against external vibration
than the STM method, it allows more precise experiments on individual
contacts. 
In both methods, the cross sectional area of the nanowire is inferred
from conductance measurements using the corrected Sharvin formula \cite{UBSG06} 
\begin{equation}
\label{eq:sharvin}
\frac{G_s}{G_0} = \frac{k_F^{2}{\mathcal A}}{4\pi} - \frac{k_F {\mathcal P}}{4\pi} + \frac{1}{6},
\end{equation}
which gives an approximation $G_s$ to the quantized conductance of an ideal metal nanowire 
in terms of geometrical quantities 
such as the wire's minimal cross-sectional area ${\mathcal A}$ and corresponding cross section perimeter
${\mathcal P}$. Here $G_0=\frac{2e^2}{h}$ is the quantum of conductance and $k_F$ is the Fermi wavevector of the material.

The formation and breakup process
is repeated thousands of times to derive a statistical histogram for
the conductance~\cite{Oleson95,CGO97,YYR99} (and therefore
cross-sectional areas). The experiments can be performed at either
ambient or cryogenic temperatures. 
Very
small contacts consisting of four gold atoms in a row have been formed
by means of this technique~\cite{OKT98}.
Although this type of experiment does not measure lifetimes of nanowires directly, rough estimates 
can be inferred from the existence of a conductance peak (which is evidence of a more stable wire, see Ref.\ \cite{Mares07}) 
by knowing parameters of the experiment's dynamics, in particular the speed of elongation of the wire.
Typically, the existence of a conductance peak in a MCBJ experiment \cite{YYR99} implies a nanowire lifetime greater than one millisecond.

Based on these criteria, there is ample experimental evidence that
nanowires made from sodium, gold 
and aluminum 
are stable with lifetimes greater than or of order milliseconds
\cite{Urban04,UBSG06,Mares07}.

\begin{figure}[htb]
  \begin{center}
	{\it $V_{shell}$}\\
	\includegraphics[angle=0,width=1.0\columnwidth]{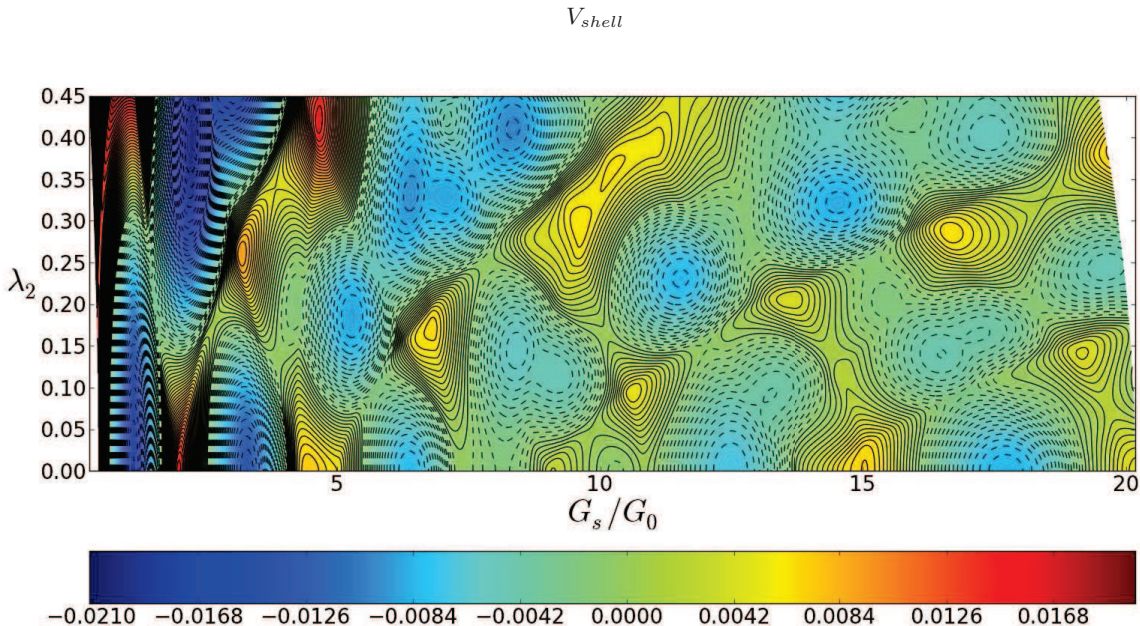}
  \end{center}
\caption{Contour plot of the electron-shell potential $V_{shell}$ as a function of 
the Sharvin conductance $G_s$ defined by Eq.\ (\ref{eq:sharvin}) 
and the quadrupolar deformation parameter $\lambda_2$.}
\label{fig:vshell}
\end{figure}

From a theoretical point of view, the stability of metal nanowires
can be explained by quantum-size effects, or electron-shell effects, 
that can overcome the classical Rayleigh instability \cite{Plateau1873} in very thin wires. 
While the Rayleigh instability makes thin wires unstable once 
their surface tension exceeds their yield strength \cite{Zhang03}, shell effects can stabilize wires with certain preferred cross sections. 
The nanowire is a quantum system where conduction electrons are confined within the
surface of the wire, with a Fermi wavelength comparable to the cross section linear dimension.
This leads to electron-shell filling \cite{KSBG99, SBB97, RDEU97}, which provides
an oscillating potential with multiple minima as a function of the cross section size and shape.
This potential is called the {\it electron-shell potential} and its derivation will be discussed in detail in
Sec.~\ref{sec:model}. Figure~\ref{fig:vshell} shows this shell potential for wires with a quadrupolar cross section as a function
of the wire's Sharvin conductance $G_s$ (related to the cross section area through Eq.~(\ref{eq:sharvin})) 
and parameter $\lambda_2$ which describes the cross-section's deviation from a disk (with $\lambda_2=0$).

Quantum effects can thus stabilize the wire against breakup due to the
Rayleigh instability. A linear stability analysis~\cite{UBSG06,KSGG01,UG03} shows that cylindrical wires are most stable
when the radius of the cross section takes a series of discrete values
called the ``magic radii". 
It also finds that some wires with non-axisymmetric cross sections can be stabilized
by electron-shell effects. 
In fact, comparison of the sequence of stable structures observed in experiments with
linear stability analyses implies that some of the experimentally observed conductance
peaks must correspond to wires with nonaxisymmetric cross sections~\cite{Urban04,Mares07}.

However, linear stability analyses typically consider only stability towards small, long-wavelength
deformations of the wire (see Refs.\ \cite{UG03,USG07} for a linear analysis including short-wavelength deformations). In particular, they do not consider 
changes of radius or breakup due to large fluctuations, which can be
initiated by thermal noise, application of stress, or other
destabilizing effects.  Because 
the regions of stability are confined by finite energy barriers
(as can be seen from Fig.~\ref{fig:vshell}), a given structure
is at best metastable; its confining barriers can be surmounted
by thermal (or other) noise. 

The energy contours shown in Fig.~\ref{fig:vshell} lead 
naturally to a description 
of the lifetime problem
in which (meta)stable structures correspond to
local energy minima; in the limit of low noise (thermal energy small
compared to the lowest confining barrier), there is an {\it optimal
  transition path} along which the probability of a successful
transition is maximized~\cite{HTB90}. When the zero-noise dynamics are
governed by a potential function (as is the case here), the
``barrier'' corresponds to a saddle point in the potential surface,
that is, a fixed point whose linearized dynamics has a single unstable
direction, all the others being stable. The transition rate is
governed by Kramers' formula
\begin{gather}
\label{eq:rate}
\mathit
\Gamma \sim \Gamma_0 \, \exp(-\Delta E/k_B T)
\end{gather}
where $\Delta E$ is the difference in energy between the saddle and
the initial state and $\Gamma_0$ is a prefactor depending on the
fluctuations around each. In the low-noise limit, the lifetime of a
metastable state is just the inverse of the Kramers' transition rate.
The problem then reduces to finding $\Delta E$ and $\Gamma_0$ as a
function of the system parameters.

In the limit of weak noise, Wentzell and Freidlin (WF)~\cite{VF}
showed that a rigorous asymptotic estimate for the rate of transition
(i.e., probability of a successful transition per unit time) can be
found by constructing an {\it action functional}, which gives the
relative probabilities of different paths. It has been shown for
gradient systems~\cite{MS93a,Marder96} 
that the optimal transition path
is one in which the system climbs uphill against the gradient through
a saddle state and then relaxes toward its final state.

Solving for the transition rate therefore requires knowledge of the
saddle state. 
The model developed in~\cite{GS10} allows for an
analytical solution to the saddle state in the case of certain special
potentials. To solve the nanowire stability problem, however, we need to find the transition
path and lifetime in more general cases. The {\it string
  method}~\cite{ERV02a,ERenEric07} is a numerical scheme 
designed for this kind of problem. The procedure it uses is to first
guess the optimal path and then let it evolve freely along the
direction of steepest descent until equilibrium is reached. Details of
the application of the string method to the kind of problem discussed
here are given in~\cite{GS11}.

In the case of cylindrical wires, the cross section of the wire can in
principle shrink or grow under the influence of noise. However,
nanowires studied in experiments are typically suspended between two
electrodes that apply a strain on the wire that tends to pull it
apart; as a consequence, transitions are biased toward smaller
radii. In either case, the electrodes act as a ``particle bath'' that
can supply or remove atoms from the wire.  

The case of transitions between different radii under the assumption
of constant axisymmetry was studied in~\cite{BSS05}. This corresponds
to transitions between energy minima along the $\lambda_2=0$ axis in
Fig.~\ref{fig:vshell}, and the calculated
lifetimes of cylindrical nanowires are comparable with experimental values. 
Here we include the effects of broken axial symmetry, so that
transitions between any two minima in Fig.~\ref{fig:vshell} can occur in principle.

The space of possible transitions is large.
However, the study can be narrowed to those of greatest physical
significance, guided by linear stability analyses \cite{UBZSG04,UBSG06} and experiments on alkali metal 
nanowires \cite{YYR99,YYR00,Urban04}.
In particular, nanowires with electrical conductance $G/G_0 = 5,
9$, and $29$, where $G_0 = 2e^2/h$ is the conductance quantum, were identified as the most stable nanowire structures with broken axial symmetry.
We
therefore concentrate on transitions from these local minima. 

We will denote non-axisymmetric 
structures by $D$ (for deformed) and the cylindrical ones by $C$. So, for example, the
non-axisymmetric structure with $G/G_0 = 5$ will be denoted $D5$, the
cylindrical structure with $G/G_0 = 3$ will be denoted $C3$, and so on.

\section{The model}
\label{sec:model}
Metal nanowires have two main components which require different treatments: 
Conduction electrons have a wavelength at the Fermi surface that is of the order of 
the linear dimension of the wire cross section, and must therefore be treated quantum
mechanically~\cite{SBB97}, and positive ions which are much heavier and therefore have 
much shorter wavelengths. As a result, the ions can be treated classically \cite{Zhang03}
and have a dynamics slow compared to that of the electrons, which can be treated separately~\cite{Burki03}. 
This separation of timescales
allows for a Born-Oppenheimer approximation where conduction electrons are considered at
all times to be in equilibrium with the instantaneous ionic structure which confines them within
the wire. Furthermore, for wires in the size regime dominated by electron-shell effects, the discrete atomic structure is unimportant and can 
be replaced by a continuum of positive charge (Jellium model~\cite{Brack93,SBB97,Burki03}):
Electron shell effects are dominant over atomic shell effects in small wires
(at least up to about $40\,G_0$ for alkali metal \cite{Urban04} and Al wires~\cite{Mares07}, and are still present above that limit.)

These observations form the basis for the nanoscale free-electron model (NFEM \cite{SBB97,UBSG06,Urban10}) 
which considers electrons in the wire to be free (other than being confined within the wire) and non-interacting.
The former works best for s-shell metals, such as alkali and to some extent noble metals like gold, but has also been 
shown to perform well for metals whose Fermi surface in the extended zone scheme is nearly spherical, such as Al \cite{Mares07}.
The limitation to non-interacting electrons has been shown to be a reasonable approximation for most metal nanowires~\cite{KSG99}.

\subsection{Energetics of the nanowire}

The surface of a nanowire aligned along the $z$-axis can be described by a generalized radius 
function $R(z,\theta)$ in cylindrical coordinates, which can be written as a multipole expansion
\begin{equation}
\label{eq:deformation}
R(z,\theta)=\rho(z)\left(\sqrt{1-\sum_m\lambda_m(z)^2/2}+\sum_m\lambda_m(z)\cos[m(\theta-\theta_m(z))]\right)\!,
\end{equation} 
where the sums run over the positive integers.
$\rho(z)$ defines the mean radius at position $z$ along the wire, 
$\lambda_m(z)$ describes a multipolar deformation of order $m$, and $\theta_m(z)$ allows for a ``twisting"
of the wire cross section along $z$, which has no effect on the wire energy given the use of the adiabatic 
approximation (see below), and will therefore be dropped.
The square root in Eq.\ (\ref{eq:deformation}) has been chosen so that the cross section area $\mathcal A = \pi \rho^2(z)$.

\begin{figure}[!htp]
\begin{center}
\includegraphics[width=0.5\linewidth]{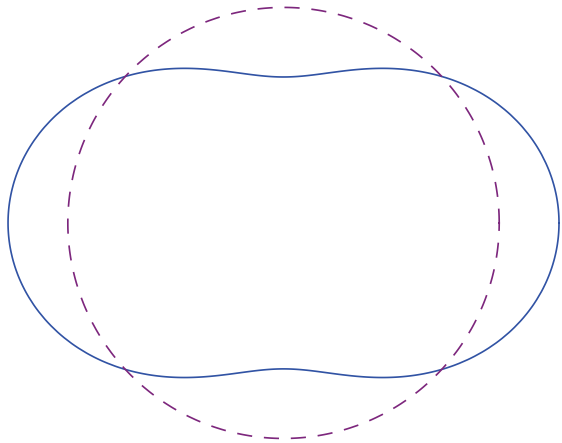}
\end{center}
\caption[Deformed cross section of a nanowire]{
Cross section (solid line) of a nanowire with a quadrupolar deformation ($m=2$).}
\label{fig:crosssection}
\end{figure}

Urban et al.~\cite{UBZSG04,UBSG06} have shown that, aside from axisymmetric wires, by far the most common stable nanowires are 
wires with a quadrupolar cross section ($m=2$). This is related to the fact that the surface-energy cost of deformations is 
proportional to $m^2$. Note that $m=1$ deformations correspond to a simple translation combined with higher-order deformations.
For that reason, we will restrict ourselves to quadrupolar deformations with $m=2$, 
so that the shape of the wire will be described by two parameters: the mean radius $\rho(z)$ and the cross section
deformation parameter $\lambda_2(z)$, which correspond to the shape shown in Fig.~\ref{fig:crosssection}.

As long as the variation of the wire cross section along $z$ is slow enough (adiabatic approximation), any thermodynamic quantity can be
written as an integral over $z$ of a local quantity that depends only on the cross section at $z$. 
This is in particular true~\cite{SBB97} for the grand-canonical potential $\Omega_e$ of the conduction electrons, which is the appropriate 
thermodynamic potential for electrons in a wire connected to bulk electrodes (open system).
$\Omega_e$ can thus be calculated for any wire of length $L$ as a functional of $\rho(z)$ and $\lambda_2(z)$
\begin{equation}
  \Omega_e[T;\rho(z),\lambda_2(z)] = \int_0^L\!\text{d}z\,\omega(T,\rho(z),\lambda_2(z)),
\end{equation}
where the local energy density $\omega(T,z)$ can be obtained numerically from the transverse energy levels $E_n(\rho(z),\lambda_2(z))$
 using the WKB approximation. The energy levels themselves can be calculated numerically for any value of $\lambda_2$ 
 and depend on $\rho$ in a simple way~\cite{SBB97,UBSG06}.
 
 On the other hand, any extensive thermodynamic quantity can be expressed in terms of a Weyl expansion~\cite{Brack97}
 \begin{equation}
\label{eq:e_grand_weyl}
\mathit
\Omega_e[\rho,\lambda_2] = -\omega \mathcal{V} + \sigma_s \mathcal{S} + \int_{0}^{L} V_{shell}[\rho,\lambda_2]\,dz,
\end{equation}
where $\mathcal{V}$ and $\mathcal{S}$ are the wire's volume and surface area, respectively.
The last term is a quantum correction and can be taken as the definition of the electron shell potential $V_{shell}$, 
depicted in Fig.~\ref{fig:vshell}.

In the spirit of the Born-Oppenheimer approximation, the electronic grand canonical
potential $\Omega_e$ is treated as the potential energy of the ions.
Since the wire can also exchange ions with the bulk electrodes, the appropriate free energy determining the structure of the wire
is the ionic grand canonical potential
\begin{equation}
\label{eq:ionic1}
\mathit
\Omega_a[\rho,\lambda_m] = \Omega_e[\rho,\lambda_m] - \mu_a {\mathcal N}_a.
\end{equation}
where $\mu_a$ is the chemical potential \cite{Burki03,BSS05} of the ions determined by the electrodes,
and ${\mathcal N}_a$ is the number of ions in the wire.

\subsection{Effective energy of deformations}

While linear stability is a necessary condition for a nanostructure to be observed experimentally, it is not sufficient
due to large stochastic deviations which can bring the wire out of its linearly stable state. Under the 
framework of the NFEM, we study the noise-induced fluctuations of the cross section by introducing
two classical fields as perturbations to the parameters $\bar{\rho}$ and $\bar{\lambda}_2$ of the generalized
radius function in Eq.\ (\ref{eq:deformation}):
\begin{gather}
\label{eq:perturbation}
\begin{aligned}
\rho(z) & = \bar{\rho}\,\left(1 + \phi_1(z)\right)\\
\lambda_2(z) & = \bar{\lambda}_{2} + \phi_2(z)
\end{aligned}
\end{gather}
where $(\bar{\rho},\bar{\lambda}_2)$ is the location of the local minimum of the ionic grand canonical potential.

Expanding the ionic grand canonical potential (\ref{eq:ionic1}) around $(\bar{\rho},
\bar{\lambda}_2)$ with respect to the fields $\phi_1$, $\phi_2$, and keeping terms up to quadratic order in the spatial derivatives, we get the
fluctuation energy functional: 
The ionic grand canonical potential (\ref{eq:ionic1}) becomes
\begin{eqnarray}
\begin{aligned}
\mathit
\Omega_a[\rho, \lambda_2] & = \Omega_a[\bar{\rho}\,(1+\phi_1), \bar{\lambda}_{2} + \phi_2]\\
& = L \left\{\pi \sigma_s \bar{\rho} \, f[\bar{\lambda}_{2}] + \left( 1 - \frac{\bar{\rho}}{2} \partial_{\bar{\rho}} \right) V_{shell} [\bar{\rho}, \bar{\lambda}_{2}] \right\} \\
&+ \int_{-L/2}^{L/2} {\mathcal H}[\phi_1(z),\phi_2(z)]\,dz,
\end{aligned}
\label{eq:ionic2}
\end{eqnarray}
where $2\pi \bar{\rho} f$ is the perimeter of the metastable nanowire's cross section, and $f$ may be represented by a quartic polynomial with high accuracy:
\begin{equation}
f[\lambda_2] = 1 +\frac{3\lambda_2^2}{4}-\frac{9\lambda_2^4}{32}.
\label{eq:f}
\end{equation}
${\mathcal H}$ is the energy density of fluctuations at position $z$ 
\begin{multline}
\mathit
{\mathcal H}[\phi_1(z),\phi_2(z)] = \\
\frac{1}{2} \pi \sigma_s \bar{\sigma}^3 \left( \left\{ 2 - \frac{\bar{\lambda}_2^2}{2} + \frac{71\bar{\lambda}_2^4}{16} \right\} (\phi_1^{'})^2  + 2 \bar{\lambda}_2 \left\{ 1 + \frac{\bar{\lambda}_2^2}{4} \right\} \phi_1^{'} \phi_2^{'} 
+ \left\{ 1 - \frac{5\bar{\lambda}_2^2}{4} \right\} (\phi_2^{'})^2\right)
\\
+ U[\phi_1(z),\phi_2(z)]
\label{eq:H}
\end{multline}
with the effective potential
\begin{gather}
\begin{aligned}
\mathit
U[\phi_1(z),\phi_2(z)] & = -\bar{\rho}\,\partial_{\bar{\rho}} V_{shell} [\bar{\rho}, \bar{\lambda}_{2}] \phi_1 + 
2 \pi \sigma_s \bar{\rho} \, f'[\bar{\lambda}_2]\phi_2 \\
& \quad - \bar{\rho} \left( \pi \sigma_s f[\bar{\lambda}_2] + \frac{1}{2} \partial_{\bar{\rho}} V_{shell} [\bar{\rho}, \bar{\lambda}_{2}] \right)\phi_1^{2} \\
&+ 2\pi \sigma_s \bar{\rho} \, f'[\bar{\lambda}_2]\phi_1\phi_2 + \pi \sigma_s \bar{\rho} f''[\bar{\lambda}_2] \phi_2^2 \\
& \quad + V_{shell} [\bar{\rho} (1+\phi_1), \bar{\lambda}_{2} + \phi_2] - V_{shell} [\bar{\rho}, \bar{\lambda}_{2}].
\end{aligned}
\end{gather}
Here $\sigma_s$ is the material-dependent surface tension: for gold $\sigma_s = 7.83\, (\mbox{eV}/\mbox{nm}^2)$; for sodium 
$\sigma_s = 1.39\, (\mbox{eV}/\mbox{nm}^2)$. 
$L$ is the length of the wire. The electron shell potential $V_{shell}$ has to be obtained separately from solving for the electronic energy bands in the 
transverse direction numerically \cite{UBSG06}.

\subsection{The dynamical system}
As discussed in~\cite{GS10} and~\cite{GS11}, the time evolution of the
fields under noise can be described by the coupled Langevin equations
$$\dot{\vec{\Phi}} = -\frac{\delta H[\vec{\Phi}]}{\delta \vec{\Phi}} + \sqrt{2\epsilon}\,\vec{\xi}(z,t)$$
where $\vec{\Phi} = \left( \begin{matrix} \phi_1 \\
    \phi_2 \end{matrix} \right)$ and $\vec{\xi}(z,t)$ is Gaussian
spatiotemporal white noise with $\epsilon = k_B T$. The saddle state
is an extremum of the action $H[\vec{\Phi}]$~\cite{HTB90} and so the
optimal transition path can be obtained numerically by evolving
$\vec{\Phi}$ according to the noise-free form of the dynamical
equations above.

For convenience, we make a change of variable $\phi_1, \phi_2 \to
\psi_1, \psi_2$ to eliminate the cross term $\phi_1'\phi_2'$ in
Eq.~(\ref{eq:H}). The two new fields are defined as:
\begin{gather}
\begin{aligned}
\mathit
\psi_1 & = \phi_1 + \frac{g_{12}}{g_1} \phi_2 \\
\psi_2 & = \phi_2
\end{aligned}
\intertext{with}
\begin{aligned}
g_1 & = \pi \sigma_s \bar{\rho}^3 (2 - \frac{\bar{\lambda}_2^2}{2} + \frac{71\bar{\lambda}_2^4}{16} ) \\
g_{12} & = \pi \sigma_s \bar{\rho}^3 \bar{\lambda}_2 ( 1 + \frac{\bar{\lambda}_2^2}{4} ) \\
g_2 & = \pi \sigma_s \bar{\rho}^3 ( 1 - \frac{5\bar{\lambda}_2^2}{4} )\, .
\end{aligned}
\end{gather}

The resulting equations of motion for $\psi_1$ and $\psi_2$ are
\begin{equation}
\begin{aligned}
\dot{\psi}_1 & =  g_1 \psi_1^{''} + \bar{\rho}(\partial_{\bar{\rho}}V_{shell}[\bar{\rho},\bar{\lambda}_2]) + 2 \bar{\rho} (\pi \sigma_s f[\bar{\lambda}_2] + \frac{1}{2} \partial_{\bar{\rho}} V_{shell}[\bar{\rho},\bar{\lambda}_2] ) \psi_1 \\
& \quad - (2\pi \sigma_s \bar{\rho} f'[\bar{\lambda}_2] + 2 \bar{\rho} (\pi \sigma_s f[\bar{\lambda}_2] + \frac{1}{2} \partial_{\bar{\rho}} V_{shell}[\bar{\rho},\bar{\lambda}_2]) \frac{g_{12}}{g_1} )\psi_2\\
& \quad - \bar{\rho} \, \partial_{\rho}V_{shell}[\rho,\lambda_2] \\
\dot{\psi}_2 & = (g_2 - \frac{g_{12}^2}{g_1}) \psi_2^{''} - ( 2 \pi \sigma_s \bar{\rho} f'[\bar{\lambda}_2] + 
\bar{\rho} \, \partial_{\bar{\rho}} V_{shell}[\bar{\rho},\bar{\lambda}_2]\frac{g_{12}}{g_1}) \\
& \quad + (2\bar{\rho}(\pi \sigma_s f[\bar{\lambda}_2] + \frac{1}{2} \partial_{\bar{\rho}} V_{shell}[\bar{\rho},\bar{\lambda}_2] ) \frac{g_{12}^2}{g_1^2} + \pi \sigma_s \bar{\rho} f'[\bar{\lambda}_2] \frac{g_{12}}{g_1} - 2\pi \sigma_s \bar{\rho} f^{''}[\bar{\lambda}_2]) \psi_2  \\
& \quad - (2\pi \sigma_s \bar{\rho} f'[\bar{\lambda}_2] + 2 \bar{\rho} (\pi \sigma_s f[\bar{\lambda}_2] + \frac{1}{2} \partial_{\bar{\rho}} V_{shell}[\bar{\rho},\bar{\lambda}_2]) \frac{g_{12}}{g_1} )\psi_1 \\
& \quad + \frac{g_{12}}{g_1}\bar{\rho} \, \partial_{\rho}V_{shell}[\rho,\lambda_2]- \partial_{\lambda_2}V_{shell}[\rho, \lambda_2]
\end{aligned}
\end{equation}
where $\dot{\psi}=\partial\psi/\partial t$ and
$\psi'=\partial\psi/\partial z$. We now turn to the solution of the
optimal transition path using the string method.

\section{Results}
\label{sec:results}

In~\cite{GS11} the string method was applied to the problem of
noise-induced transitions in a 
two-component classical field theory.
The
transition path, or {\em string}, starts in a random configuration on
the potential surface, with its two ends inside the basins of
attraction of the initial and final states, respectively.  The string
is then allowed to evolve along the direction of the energy gradient,
thereby determining the optimal transition path.  The saddle state is
the configuration of highest energy along this path. The results were
consistent with the analytical solutions of~\cite{GS10}, in particular,
the transition of the saddle state from a homogeneous to an instanton
configuration as $L$ increases beyond a critical value $L_c$.

In the following, we apply the same numerical scheme to the ionic grand
canonical potential surface of a metal nanowire, in order to study lifetimes of wires
whose cross sections correspond to the conductance plateaus $D5$ and $D9$.

\begin{figure}[htb]
  \begin{center}
	\includegraphics[width=0.8\linewidth]{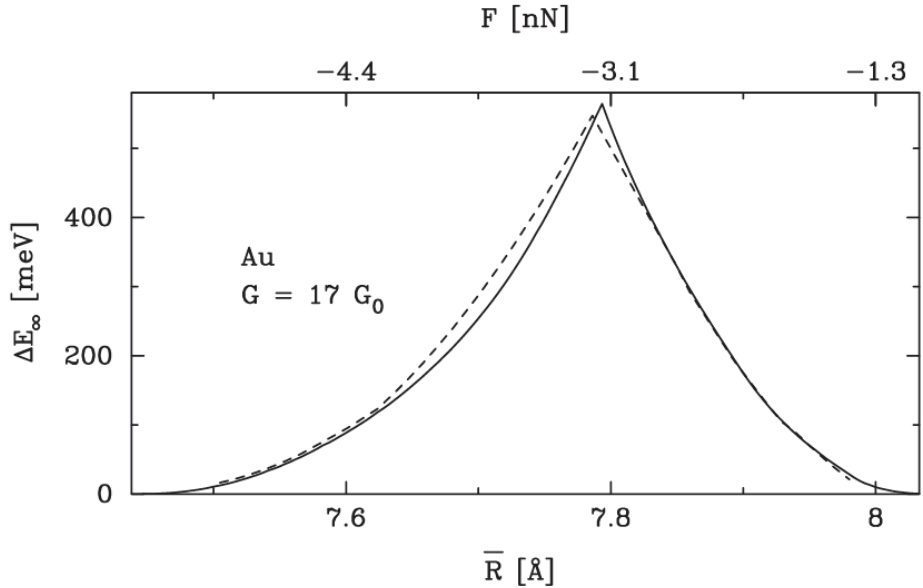}
  \end{center}
\caption[Activation barrier at infinite length vs. radius]
{The activation barrier $\Delta E_{\infty}$ as a function of mean radius for the
conductance plateau $G_s/G_0 = 17$ of gold. Solid curve: numerical result for the
full potential; dashed curve: result from theoretical
calculation~\cite{BSS05} using the best cubic-polynomial fit to the full
potential. The mean radius is related to the tensile stress (upper axis). This
figure is taken from Ref.~\cite{BSS05}.}
\label{fig:BSS05}
\end{figure}

It was found theoretically in~\cite{BSS05} that both the activation barriers and the
transition direction are sensitive to changes in stress of the order of 1nN.
Fig.~\ref{fig:BSS05} shows the activation barrier for the transition of a cylindrical gold nanowire 
on the conductance plateau $G/G_0=17$ (denoted $C17$) to other linearly stable
structures. To the left of the cusp, the transition proceeds in a direction
corresponding to thinning 
(equivalently, moving to a lower conductance value); to the right, the transition proceeds via thickening
(moving to a higher conductance value). The most stable structure of $C17$ corresponds
to the maximum value of $\Delta E_{\infty}$, which is located at the cusp;
at this point the activation barriers for thinning and growth are
equal. 

For a non-axisymmetric wire, by contrast, it is unlikely for thinning
and growth processes to reach equilibrium at the same stress; the cusp
in Fig.~\ref{fig:BSS05} is therefore absent. 
The greater richness of
the configuration space for non-axisymmetric wires allows more options for 
escape from a given initial state; barriers are therefore generally lower
(and lifetimes shorter) than in situations restricted to axial
symmetry.

\begin{figure}[h]
\centering
\includegraphics[width=\linewidth]{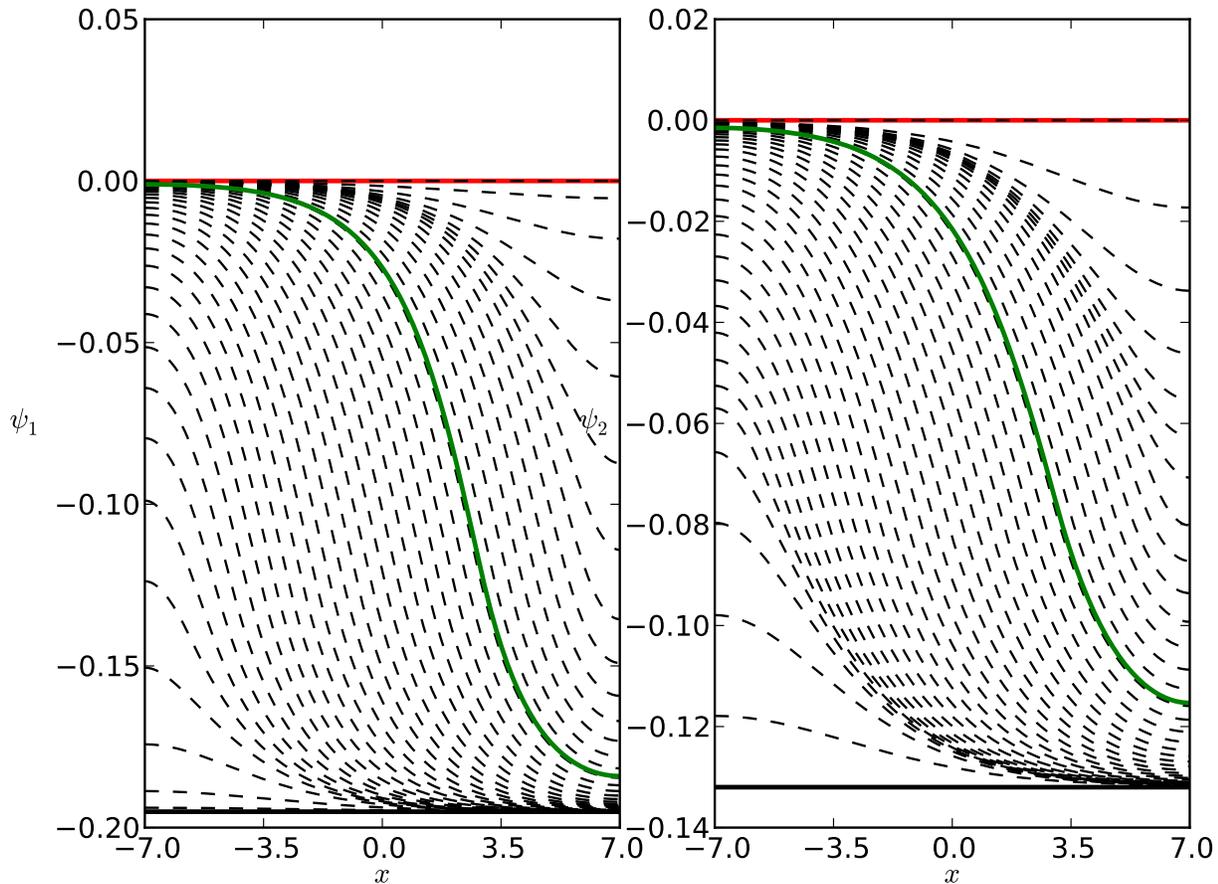}
\caption[short instanton]{A typical configuration of the short instanton (green), where the transition
starts from the initial $D5$ configuration at $G_s/G_0=5.55$ (red line) and ends at its nearest neighbor state $C3$ (thick black). 
Here a gold wire of length 1.2nm is considered. Dashed black lines are other intermediate states along the transition path.}
\label{fig:shortins}
\end{figure}

\begin{figure}[h]
\centering
\includegraphics[width=\linewidth]{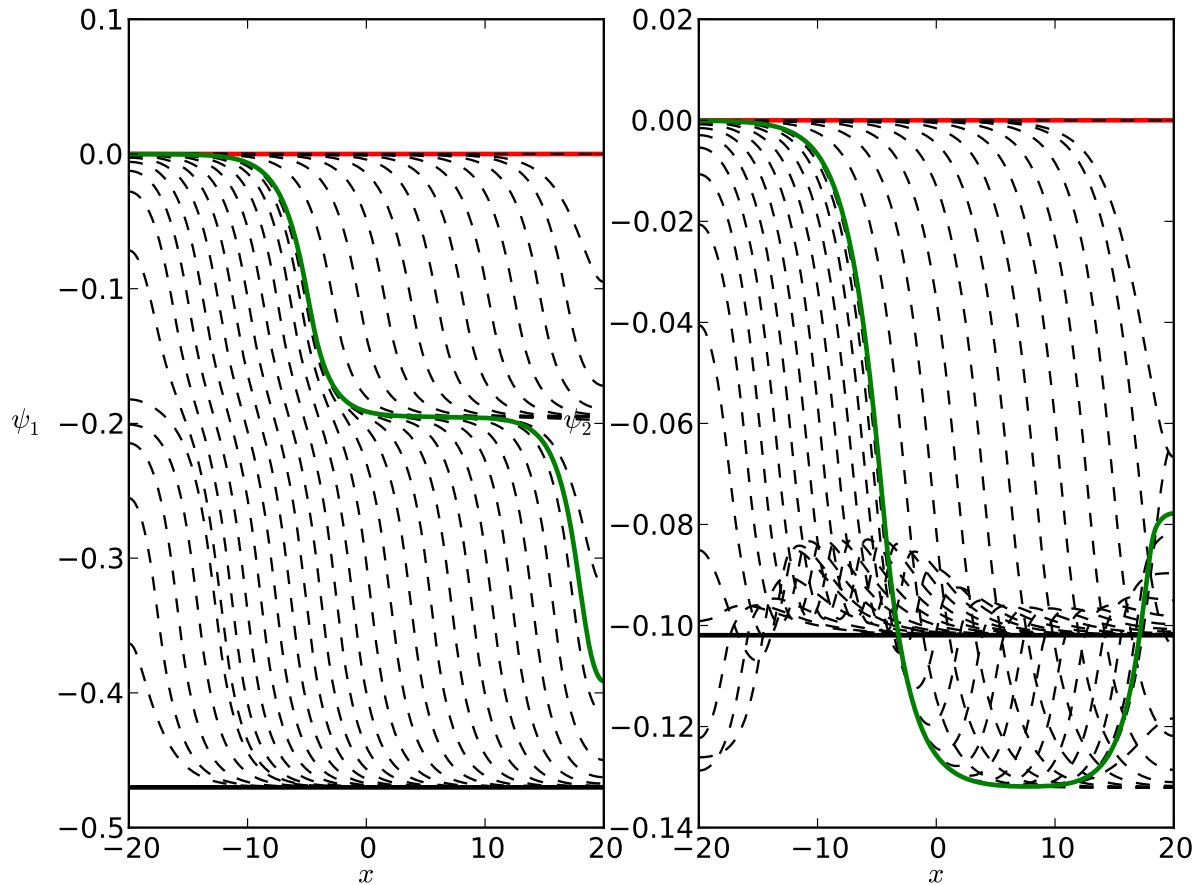}
\caption[long instanton]{A typical configuration of the long instanton, where the transition starts from the initial $D5$ configuration at
$G_s/G_0=5.56$ and ends at its second-nearest neighbor state $C1$. 
Note that there are two obvious plateaus in the shape of the long instanton, with one corresponding to the initial state and the other to the 
intermediate local minimum $C3$. Here a gold wire of length 3.3nm is considered. The line styles follow the same convention as in Fig.~\ref{fig:shortins}.}
\label{fig:longins}
\end{figure}

Detailed analysis shows that there are two types of instanton transition states that may arise as the
initial state of the wire varies. The first corresponds to decay into
the nearest cylindrical structure, and
resembles an asymmetric hyperbolic tangent function with
the longer arm coinciding with the initial configuration (see Fig.~\ref{fig:shortins}). 
The second corresponds to decay into (usually) the second-nearest neighbor
cylindrical structure, and
consists of multiple plateaus, each
corresponding to a local minimum the transition goes through (see
Fig.~\ref{fig:longins}). We hereafter refer to the first
as a ``short instanton'' and the second as a ``long instanton.''  The
final state of the latter is farther from the initial state in configuration 
space than that of the former.

\begin{figure}[h]
\centering
\includegraphics[width=\linewidth]{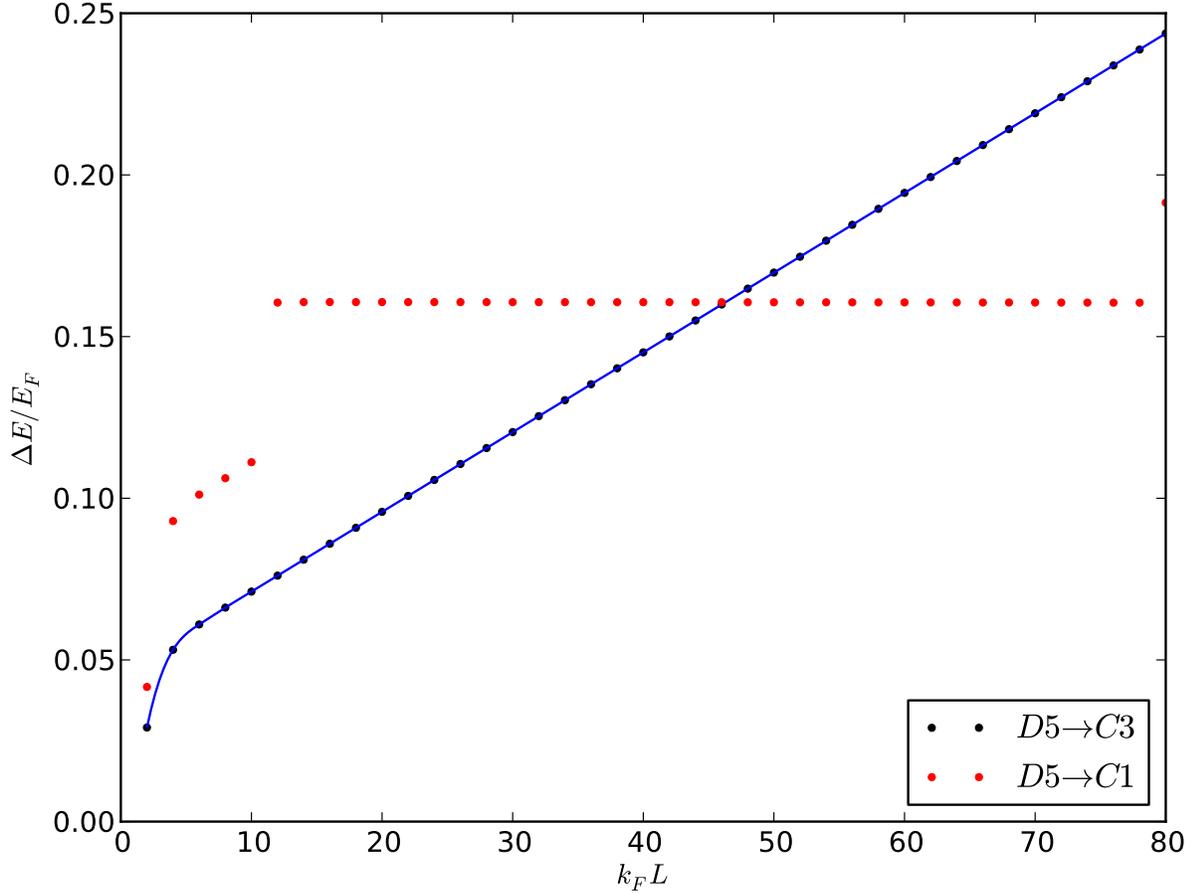}
\caption[Switch from short-instanton to long-instanton]
{Activation barrier for thinning of the sodium $D5$ state, where the initial structure has $G_s/G_0=5.52$. 
There are two different final states: $C3$ and $C1$, to which the transition is via the short  ($C3$) and long instanton ($C1$), respectively. 
The x-axis is in dimensionless units whose range corresponds to 0 to 8.8nm. 
Note that there is a discontinuity in the barrier of the transition $D5 \to C1$, below which it suddenly drops and becomes inclined. 
For very short wires, the long instanton is not favorable and ``collapses'' into two short instantons, where the first is the saddle state of the 
transition $D5 \to C3$ and the second is the saddle state of the transition $C3 \to C1$. 
The string method can only select the saddle of the highest energy if there is more than one along the transition path; 
here it is that of $C3 \to C1$, so that the activation barrier calculated increases with $L$.}
\label{fig:barrierswitch}
\end{figure}

Switching between short and long instantons is caused by the change
in the behavior of the activation barrier as a function of $L$. Within a given
family of wires (e.g., $D5$) lying in a particular
basin of attraction of the electron shell potential, there is a qualitative
difference in the activation behavior of the thinner wires versus the thicker wires, 
which can be tuned by applying tensile/compressive stress.
For thinner wires (e.g., under tensile stress), the energy of the short instanton first grows for $L<L_c$
and then reaches a plateau for $L>L_c$; for thicker wires (e.g., under compression), it
continues growing as $L$ increases beyond $L_c$. The upper bound of
the lifetime for the transition consequently grows without bound in the latter case as
$L$ increases. Conversely, the energy of the long instanton approaches
a finite asymptotic value as $L \to \infty$. Therefore, for thicker wires 
in a given family,
the energies of the short  and long instanton cross at a certain
$L$, beyond which the short instanton transition state is no longer
favorable (see Fig.~\ref{fig:barrierswitch}).

The change in the energy behavior of the short instanton can be
understood by modeling the transition using an asymmetric double well
potential.  In such a system (shown in Fig.~\ref{fig:forward}), the
activation barrier of the state belonging to the upper well is finite
and independent of $L$ as $L \to \infty$; consequently, the
leading-order exponential term determining the lifetime
[Eq.~(\ref{eq:rate})] approaches a constant value. In the other
direction, however, the lifetime of the lower well, as determined by
its transition rate to the upper one (shown in
Fig.~\ref{fig:backward}), is not bounded because the activation
barrier grows with $L$: $\Delta E \sim C \cdot L$, with $C$ a constant
determined by the energy difference between the two wells. When 
the cross section of the wire is varied by adjusting the applied tensile
force, the two potential wells (representing the two linearly stable
states) shift vertically relative to one another, until the lower well
becomes the new upper well and vice-versa. 

\begin{figure}[h]
\centering
\includegraphics[width=\textwidth]{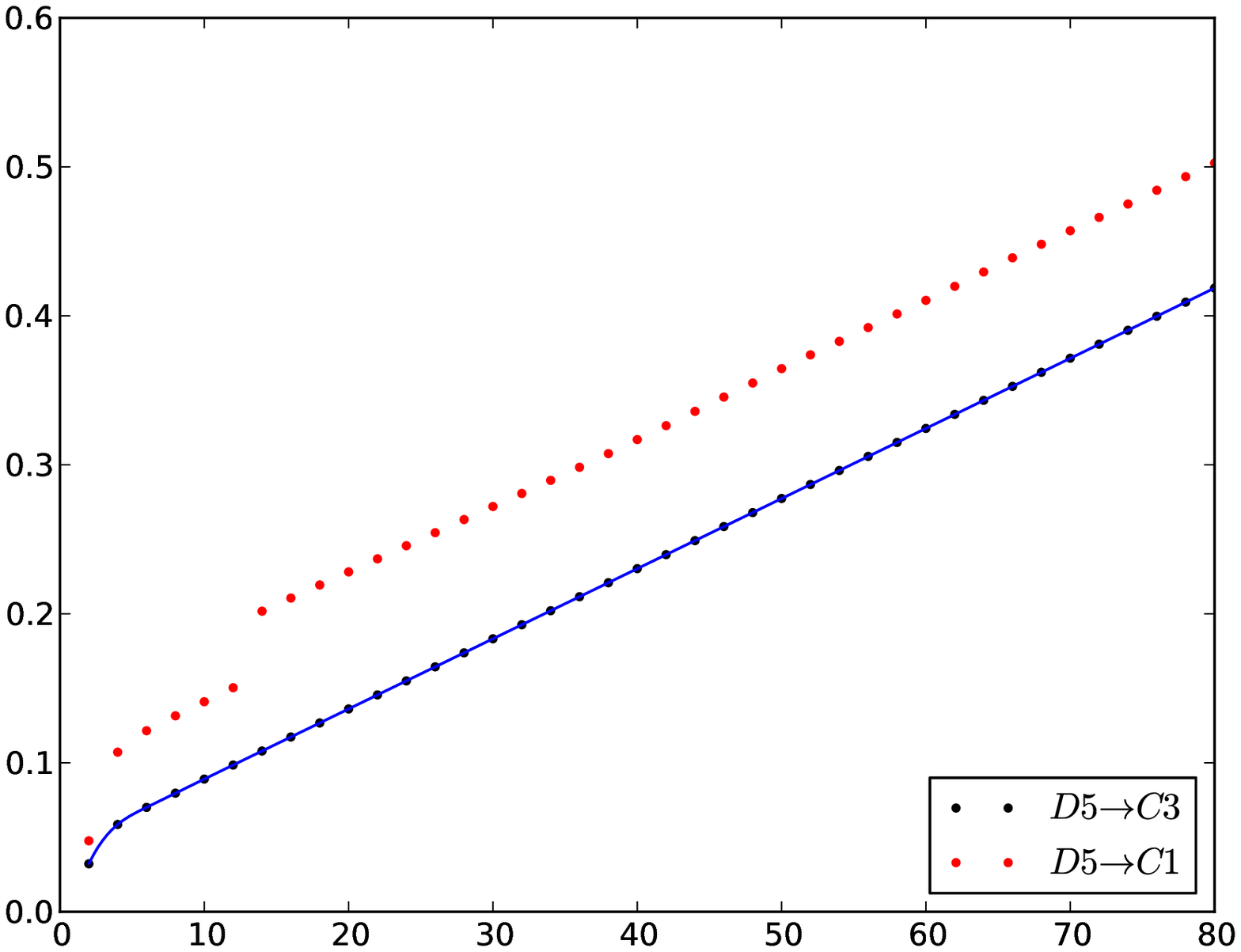}
\caption{A typical situation at $G_s/G_0 = 5.56$ where the transition via long instanton turns unfavorable following that of the short instanton. 
The activation barrier of the long instanton (red dots) increases with $L$ as well as the lifetime so that the wire becomes more and more stable against 
thinning. The transition direction switches accordingly from thinning to growth.}
\label{fig:unfavlongins}
\end{figure}

Consider now the thinning process for wires under tension; as an
example, we study the situation where the state $D5$ corresponds to
the upper well and $C3$ to the lower well. Under tension, the activation
barrier $\Delta E_{\infty}$, and hence the lifetime of the thinning
process from $D5 \to C3$, is always bounded. Under compression, however, $D5$ shifts downward to become the lower well and
$C3$ the upper; now $\Delta E_{\infty}$ and the corresponding lifetime
become unbounded as $L \to \infty$. The implication from this
asymmetric double well model is that, to have a bounded lifetime, it
is necessary to find a final state whose energy is lower than that of
$D5$. $C1$ is such a state; the lifetime of $D5 \to C1$ is bounded
while that of $D5 \to C3$ is not. These results are summarized in
Fig.~\ref{fig:contrastfb}.

We can infer some of the dynamics of the process of escape from the
shape of the long instanton. Fig.~\ref{fig:longins} implies that
during the escape process part of the wire assumes the $C3$ state,
which then bends further towards $C1$. We do not expect the long
instanton to be a relevant intermediate state for escape from short
wires as it would cost excessive bending energy in forming the
necessary critical droplet. This conclusion is consistent with
Fig.~\ref{fig:barrierswitch}, in which the flat barrier of the
long instanton disappears at small $L$.

\begin{figure}[h]
	\centering
	\subfloat[Transition pattern of $D5$]{
\label{fig:patternD5}
\includegraphics[width=0.45\linewidth, height = 0.2\textheight]{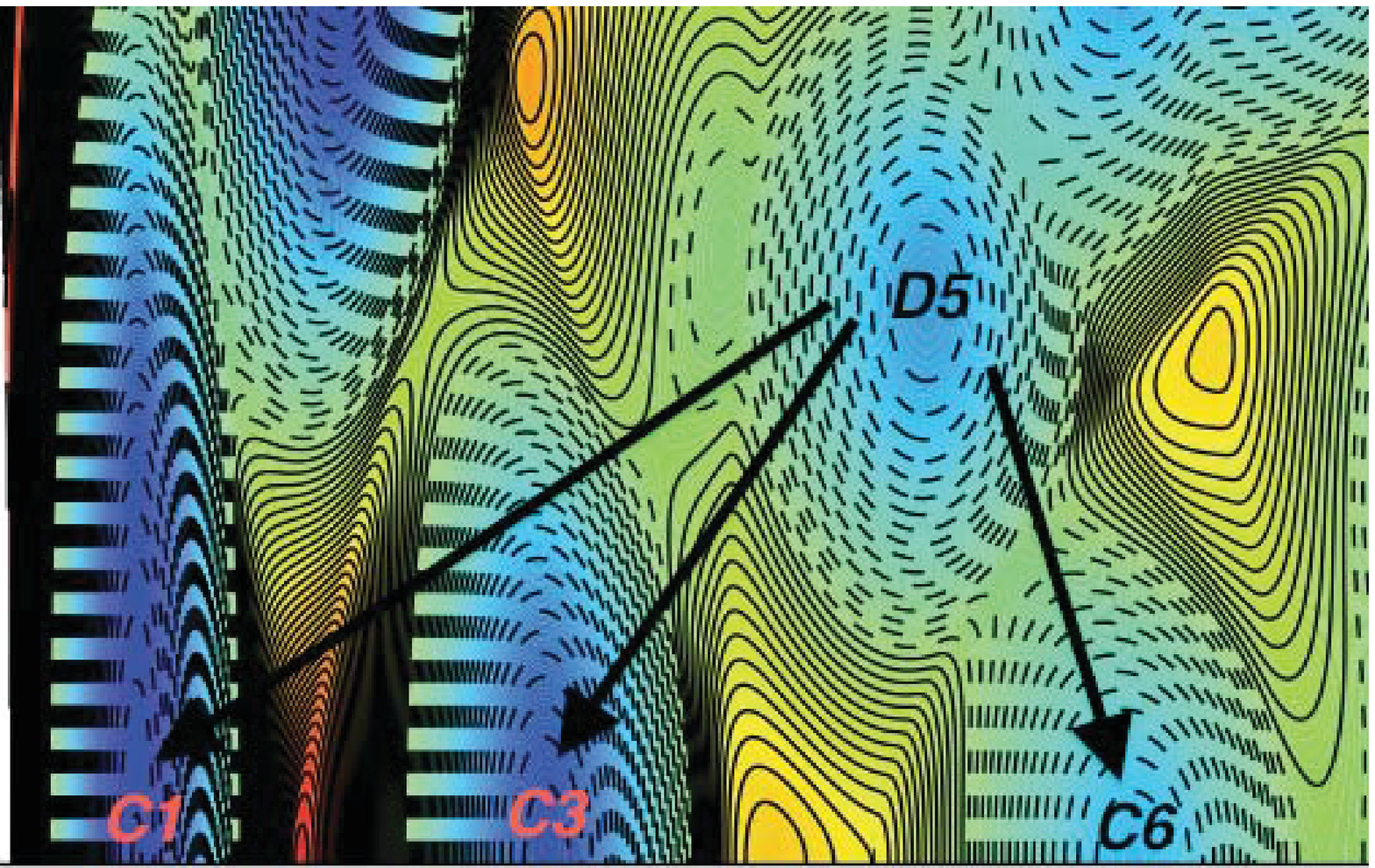}} \quad
	\subfloat[Transition pattern of $D9$]{
\label{fig:patternD9}
\includegraphics[width=0.45\linewidth, height = 0.2\textheight]{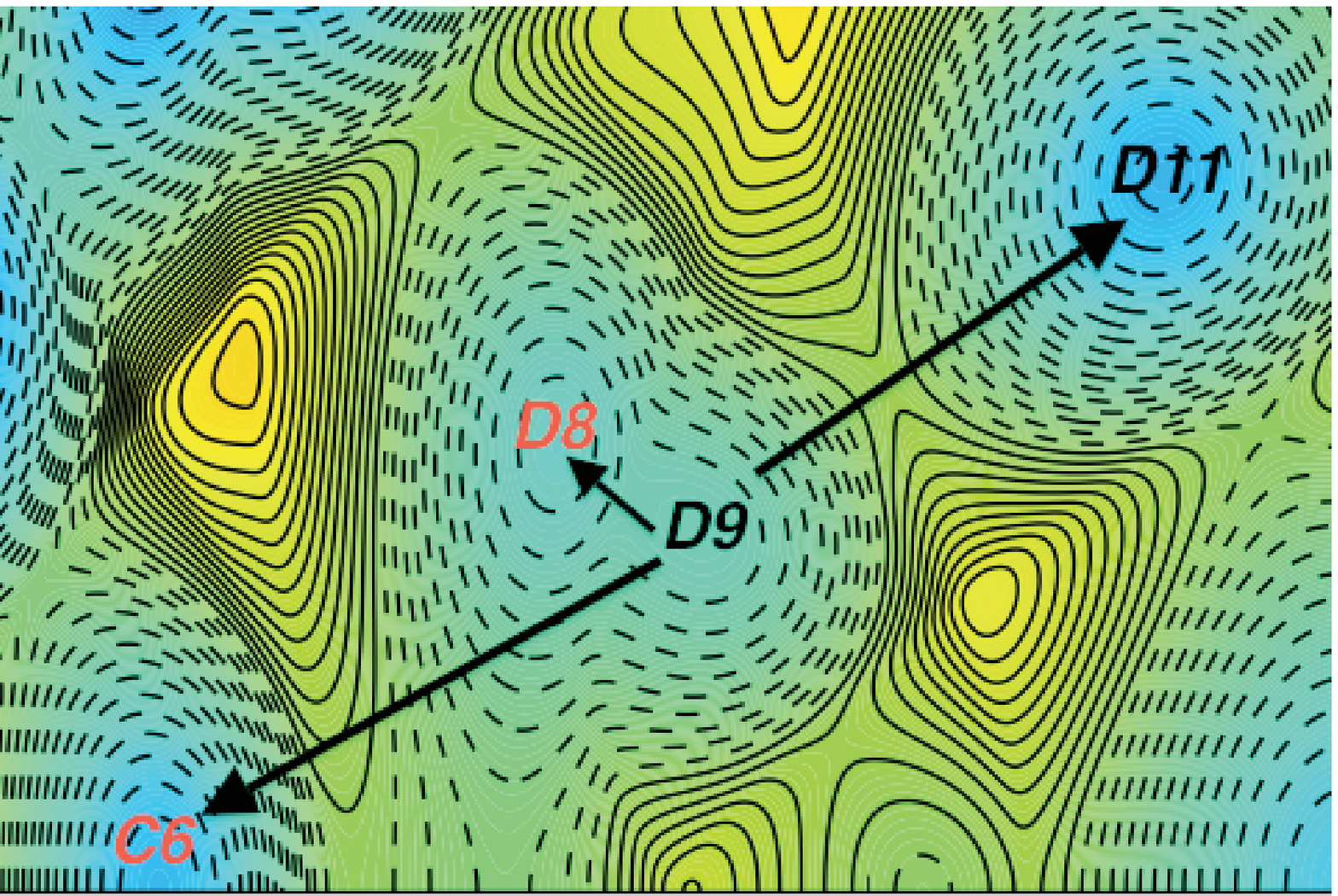}}
	\caption{Transition patterns for the structures $D5$ and $D9$. Both thinning and growth are drawn, 
where the final states of thinning are highlighted in red. Note that for growth only the nearest final state ($C6$ for $D5$ and $D11$ for $D9$) 
is shown for discussion purposes.}
\end{figure}

\begin{figure}[h]
\centering
\emph{The switch from short-instanton to long-instanton}\\
\subfloat[Forward transition via short-instanton]{
\label{fig:forward}\includegraphics[width=0.45\textwidth,height=0.2\textheight]{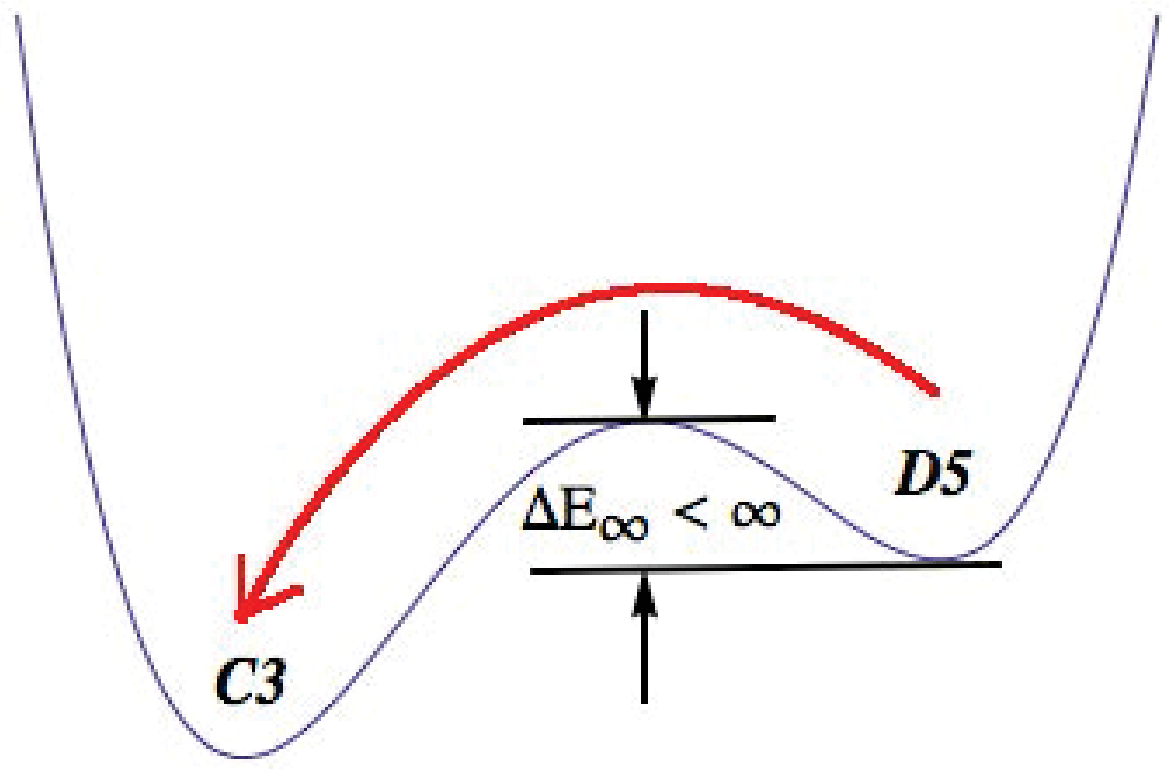}} \quad
\subfloat[Backward transition via short-instanton]{
\label{fig:backward}\includegraphics[width=0.45\textwidth,height=0.2\textheight]{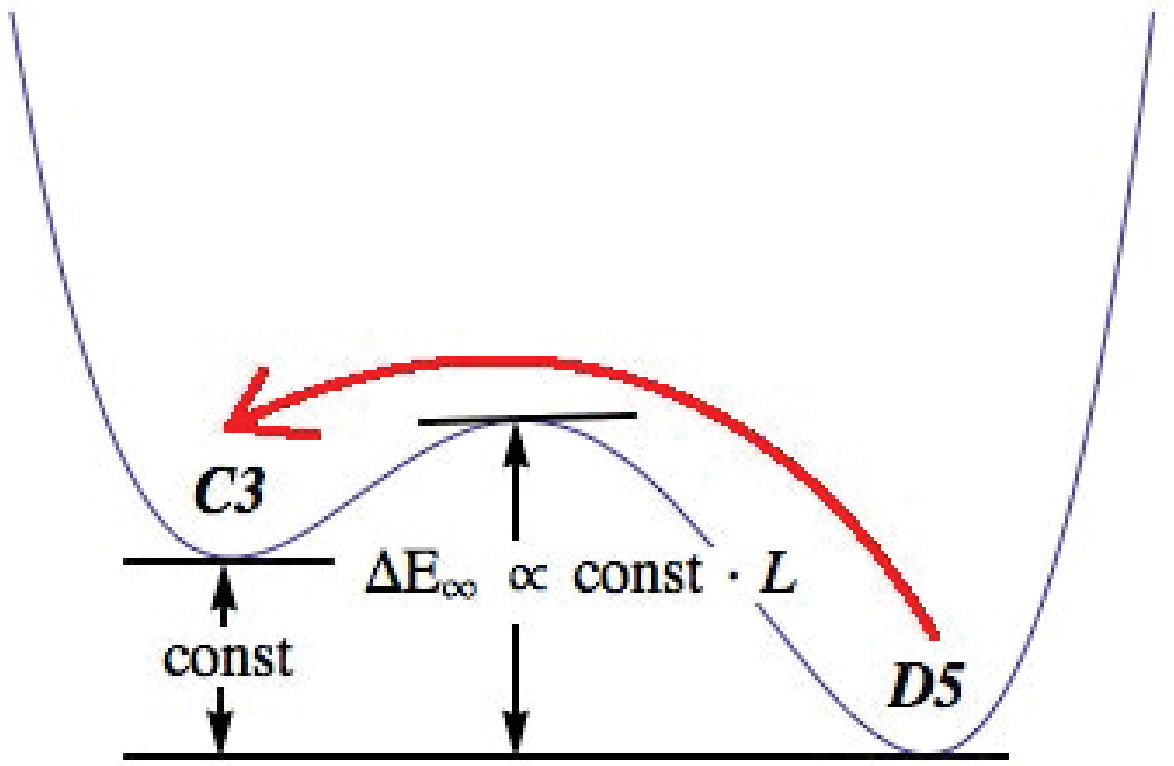}} \\
\subfloat[Barrier of forward transition]{\label{fig:energyD5toC3boundedAu}\includegraphics[width=0.45\textwidth,height=0.2\textheight]{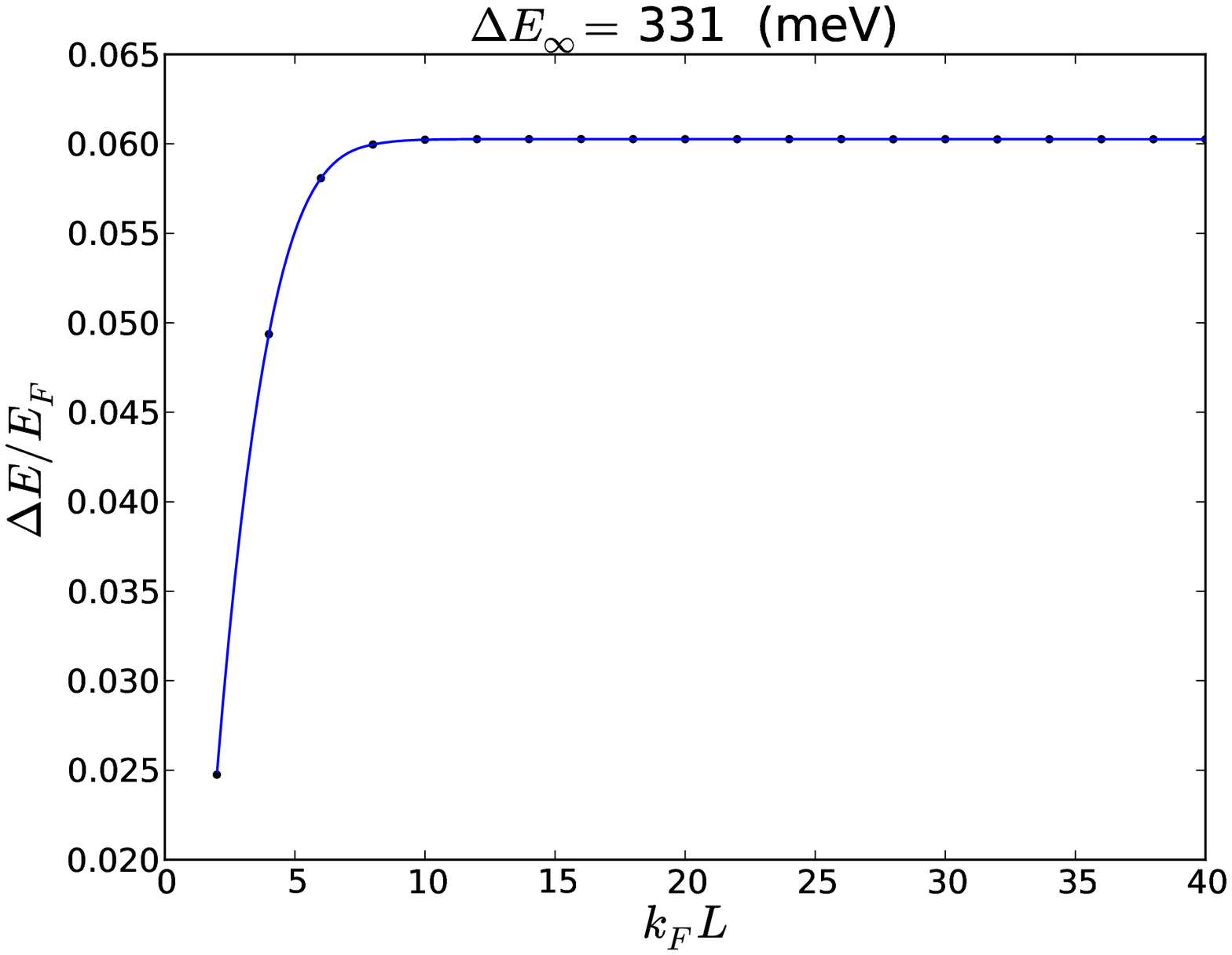}}\quad
\subfloat[Barrier of backward transition]{\label{fig:energyD5toC3unboundedAu}\includegraphics[width=0.45\textwidth,height=0.2\textheight]{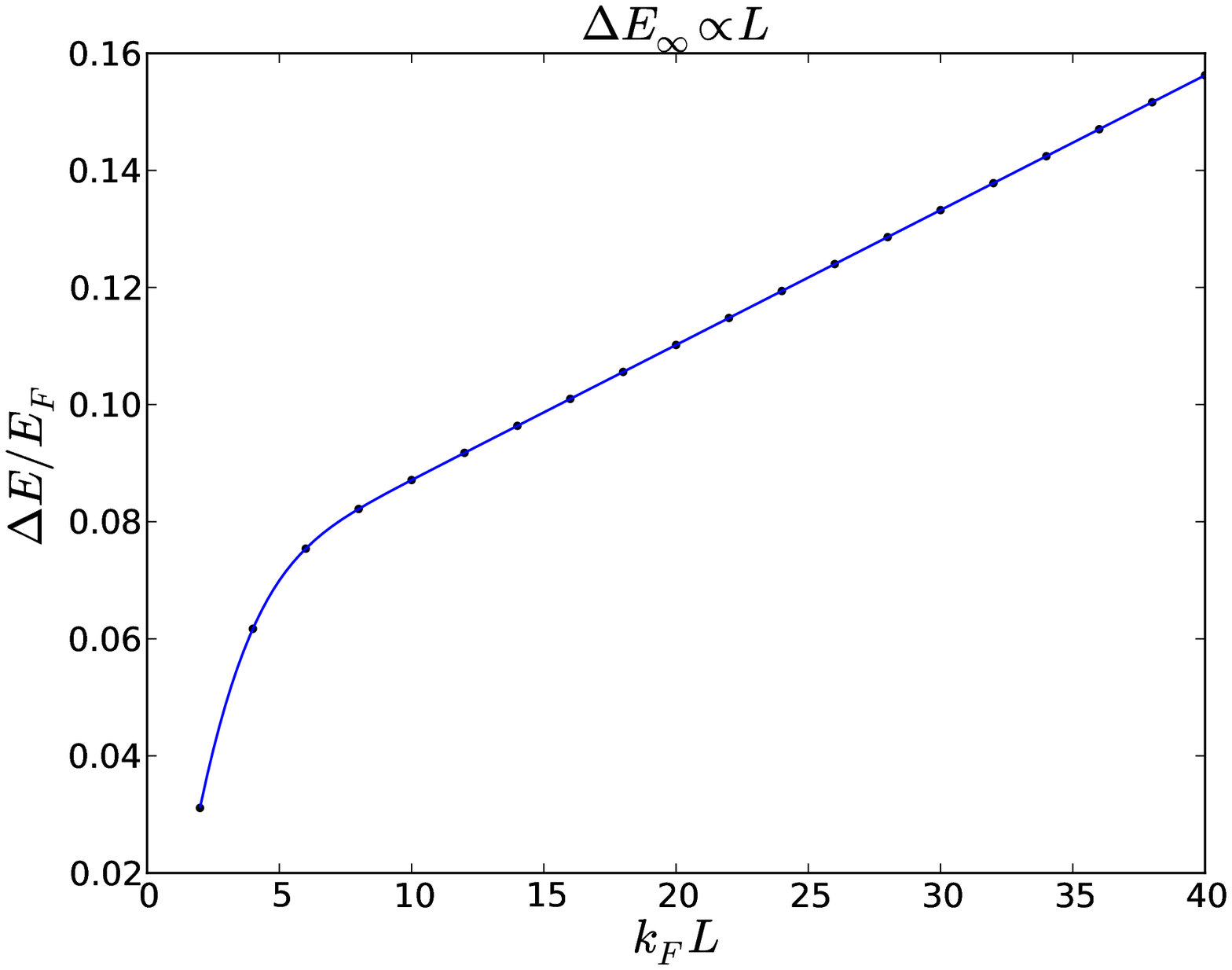}}\\
\subfloat[The onset of long instanton]{\label{fig:D5toC1}\includegraphics[width=0.45\linewidth,height=0.2\textheight]{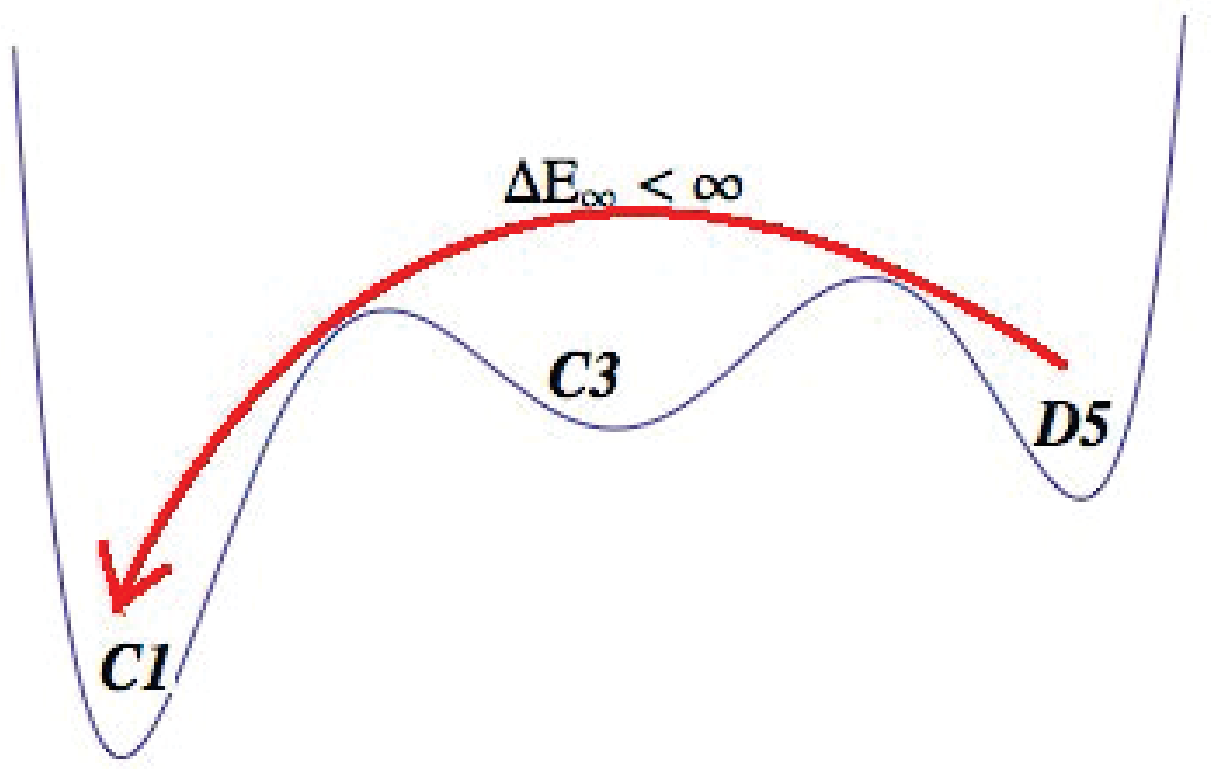}}
\caption{The first 4 subplots show the transition and the activation barrier of the $D5 \to C3$ transition for Au via the short instanton. 
Two different situations are shown: under tension at 
$G_s/G_0=5.5$ (Fig.~\ref{fig:forward}, Fig.~\ref{fig:energyD5toC3boundedAu}) the activation behavior
displays the characteristics of the forward transition, while under compression 
at $G_s/G_0=5.6$ (Fig.~\ref{fig:backward}, Fig.~\ref{fig:energyD5toC3unboundedAu}) the activation behavior displays the characteristics of the
backward transition. The activation barrier of the forward transition levels
out as $L \to \infty$ while that of the backward transition grows almost linearly with $L$,
following the discussion of the asymmetric double well model. The greater the
difference in energy between the two metastable states, the faster $\Delta E$
grows with $L$. The last subplot sketches the transition from $D5$ to $C1$ via the long instanton, which sets a bound
(as shown in Fig.~(\ref{fig:barrierswitch})) on the barrier for thinning of $D5$ when the transition to $C3$ is backward.}
\label{fig:contrastfb}
\end{figure}

As the wire is compressed and its cross section continues to grow, the energy of even
$C1$ starts to shift upward and eventually it becomes the upper well
relative to $D5$; the probability of thinning thereafter decreases
(see Fig.~\ref{fig:unfavlongins}).
A similar analysis can be applied to $D9$, where the role of $C3$ is
replaced by $D8$ and $C1$ by $C6$.  The transition patterns of $D5$
and $D9$ are sketched in Fig.~\ref{fig:patternD5} and
Fig.~\ref{fig:patternD9}.

In Fig.~\ref{fig:contrastfb}, we denote the direction from the upper
well to the lower one as the ``forward'' transition and the reverse as
``backward''.  With its activation barrier reaching a plateau as $L
\to \infty$, the forward transition should occur more frequently than
the backward one; we therefore refer to it as the {\it transition
  direction}. We emphasize that the terms ``forward'' and ``backward''
are defined relative to the upper well, that is, the higher-energy
metastable state.

In experiments, the wire is constantly subject to breakup and re-formation. 
Under elongation,
the cross section typically samples all possible stable values before breakup. 
In reforming the nanowire, on the other hand, there is typically a jump to contact, and not all structures are clearly resolved in the conductance
histogram.  The relevant lifetime determining whether a given metastable structure is observed experimentally is therefore that 
governing the thinning process when the nanowire is under tension.

In the following sections, we calculate and display the lifetimes of the
$D5$ and $D9$ structures for gold and sodium nanowires. 

\subsection{Gold nanowires}
\label{subsec: results_gold}

In this subsection, we discuss our results for the stability of the
$D5$ and $D9$ structures for gold nanowires. 

\begin{figure}[h]
\centering
\includegraphics[width=\linewidth]{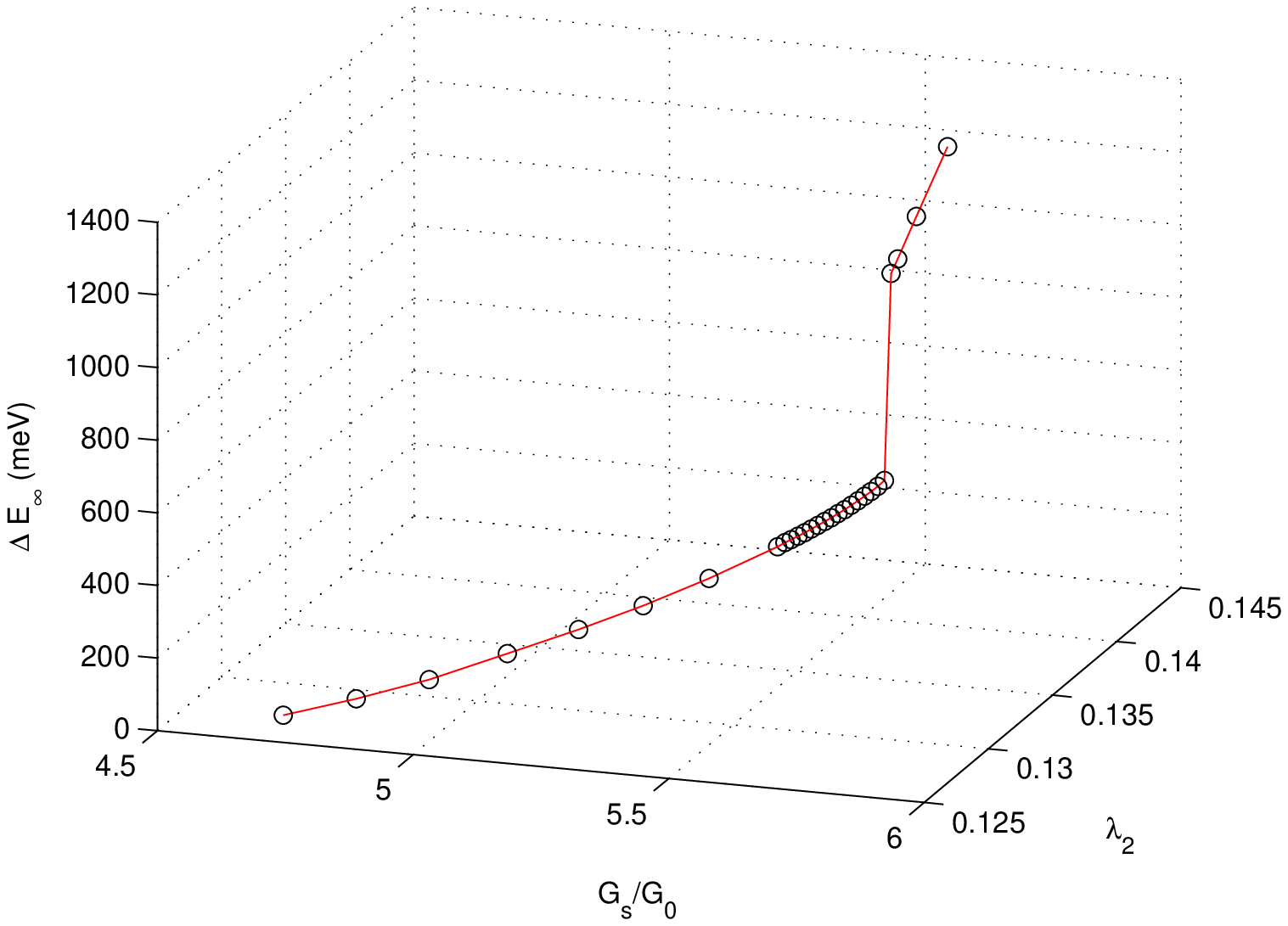}
\caption{Activation barrier for thinning of the $D5$ structure in Au. 
The $G_s/G_0$ axis is related by the Sharvin formula to the continuous variation of the cross section resulting from the change of tensile force; 
the $\lambda_2$ axis is the corresponding variation in the degree of axial symmetry breaking. 
Below the critical cross section at $G_s/G_0 = 5.56$, the activation barrier $\Delta E_{\infty}$ is determined by the transition $D5 \to C3$ via the 
short instanton, while for $G_s/G_0 > 5.56$ the activation barrier is determined by the transition $D5 \to C1$ (long instanton). The curve stops at 
the point where the energy of the long instanton becomes unbounded and thinning thus surrenders to growth.}
\label{fig:AuD5BarrierCurve}
\end{figure}

For $D5$, $G_s/G_0 = 5.56$ corresponds to the critical cross section 
beyond
which the long instanton plays a part in the determination of the
lifetime. As the wire becomes thicker, the energy of the long
instanton, and therefore the lifetime, increases
monotonically. Eventually, the transition $D5 \to C1$, occurring via
the long instanton, becomes backward, essentially halting the thinning
process. The evolution of the barrier for thinning is plotted in
Fig.~\ref{fig:AuD5BarrierCurve}; the energy of the long instanton
ranges up to $\Delta E_{\infty}=1000\,\mbox{meV}$, corresponding to a lifetime
$\tau \approx 10^{7} \mbox{sec}$ at $T = 300K$.

If one considers growth processes, the lifetime may be much
shorter. The lifetime of the $D5 \to C6$ process, for example, where $C6$ is a
cylindrical state with a higher conductance value, is only of order
$10^{-12} \mbox{sec}$. 
However, in experimental situations in which tensile stresses
bias the system toward thinning,
$D5$ should live long enough to be observed experimentally.  
Atoms can always leave the wire by
diffusing out onto the surface of the bulk electrodes, while atoms incorporated into the bulk electrodes cannot readilly migrate into the wire, so that
thickening may only be possible on reasonable timescales under compression.

For the $D9$ structure, the analysis is similar to that for $D5$, except that the
critical cross section for the onset of the long instanton is
approximately $G_s/G_0 = 9.3$. Below this value, the likeliest
transition is from $D9$ to $D8$ via the short instanton; above this value, the transition 
will be from $D9$ to $C6$ via the long instanton. For thicker wires,
$D9$ could transition to $D11$ rapidly, with a lifetime of order a
millisecond. Overall, $D9$ is quite stable against thinning, with a
lifetime (determined by the long instanton) of $10^3 \, \mbox{sec}$ at $T =
100K$. 

\subsection{Sodium nanowires}
\label{subsec: results_sodium}

The qualitative features of the stability of sodium nanowires are 
similar to those of gold nanowires. However, lifetimes are significantly shorter
(cf.\ Ref.\ \cite{BSS05}) because of sodium's smaller surface energy.

Nevertheless, $D5$ is quite stable against thinning, with a lifetime of
$0.3 \,\mbox{ms}$ at $T=100K$. 
The critical cross section above which the
long instanton becomes relevant is $G_s/G_0 = 5.48$. $D9$ has a
lifetime of $0.057\,\mbox{sec}$ at $T=60K$ or $0.00012\,\mbox{sec}$ at $T=80K$. The
critical cross section for $D9$ is $G_s/G_0=9.3$. Both plateaus are
unstable at room temperature and the same issues concerning growth
vs.\ thinning apply to sodium as for gold.

\section{Discussion}
\label{sec:discussion}

The previous section discusses a mechanism --- the interplay of
transitions via short vs.~long instantons --- that is likely to play
an important role in determining lifetimes of nonaxisymmetric wires.
Physically, this interplay reflects the competition between classical
and quantum effects (the latter encapsulated within the electron shell
potential).  In the classical regime, a long wire requires a large
cross section to stabilize against the Rayleigh instability. However,
when quantum effects are taken into account, the deep minima of the
electron shell potential favor thinner structures, giving rise to
``magic radii''. When quantum effects dominate, the thinning process
is favored and the transition from thicker to thinner is forward; when
classical effects dominate, the thinning transition becomes backward
and hence less favorable. As the cross section is continuously varied
from larger to smaller, we find that thinning is inhibited when the
initial structure is thicker; the converse when it is thinner.

In the purely cylindrical case, thinning and growth reach equilibrium
when the wire is at a particular cross-sectional area, as in Fig.~4
of Ref.\ \cite{BSS05}. The most stable structure and its lifetime can thus be
defined at the cusp.  In the non-axisymmetric case, however, there is
no similar single point of equilibrium. The rate of growth is
generally much faster than that of thinning, so that the wire may
escape to a thicker state even if the lifetime for thinning is large.
Experimentally, however, $D5$ and $D9$ are observable with lifetime of
order milliseconds or larger. We believe this is due to the
experimental situation in which applied stresses to the wire inhibit
growth.  Our model for both thinning and growth implicitly assumes
that the transfer of atoms into and out of the wire is
instantaneous. This assumption is a reasonable approximation for
thinning but not for growth under normal experimental conditions. The
observed stability of laboratory nanowires is largely restricted by
thinning processes: in experiments the wire usually either thins or
breaks up under the pulling force. Given this, our model is quite
successful in predicting the stability of $D5$ and $D9$ conductance
plateaus against thinning.

\begin{acknowledgments}
The authors are grateful to Seth Merickel for his contributions to the early stages of this work.  This work
was supported in part by NSF Grant PHY-0965015. 
C.A.S.\ acknowledges support from the U.S. Department of Energy, Basic Energy Sciences under Award No.\ DE-SC0006699. 
D.L.S.\ thanks the John Simon Guggenheim Foundation for a fellowship that partially supported this research.
\end{acknowledgments}

\bibliographystyle{ieeetr}
\bibliography{Reference}

\end{document}